\documentclass[referee]{aa}

\begin{document}

\title{Mie, Einstein and the Poynting-Robertson effect}
\author{J. Kla\v{c}ka}
\institute{Faculty of Mathematics,
Physics and Informatics, Comenius University \\
Mlynsk\'{a} dolina, 842 48 Bratislava, Slovak Republic \\
e-mail: klacka@fmph.uniba.sk}

\date{}

\abstract{
A paradox associated with the astrophysical Poynting-Robertson effect is
presented. The paradox arises when relativity theory and Mie's solution of
Maxwell's equations are confronted with the statements
on the Poynting-Robertson effect. Although the relevant
physics has been known already for a century (Poynting 1903, Einstein 1905,
Mie 1908), nobody has been aware of the inconsistency between the theories.

\keywords{Poynting-Robertson effect, electromagnetic radiation, Mie's theory,
relativity theory}
}

\maketitle

\section{Introduction}
Physicists celebrate 100 years from the Mie's paper on the scattering
of light by spherical particles (Mie 1908). Knowing the distribution of
the material characteristics (refractive index and conductivity)
within the particle, radius of the particle and wavelength(s) of the incoming
light, Mie's (and Debye's; see also van de Hulst 1981, Bohren and Huffman 1983)
calculation provides such physical quantities as cross sections of extinction,
scattering and absorption of light.
Moreover, since electromagnetic radiation generates a pressure
force, Mie's solution of Maxwell's equations for electromagnetic field
enables us to calculate cross section for radiation pressure.

Experimental evidence of the light pressure was presented by Russian
physicist P. N. Lebedev at the end of the 19-th century. This result
motivated another prominent physicist, J. M. Poynting, to formulate
the problem of motion of a perfectly absorbing spherical particle under
the action of incident light (Poynting 1903). Several different solutions
were offered afterwards, until Robertson (1937) proposed
a relativistically covariant equation of motion of the particle.
Poynting and Robertson assumed that particle's "own radiation outwards 
being equal in all directions has zero resultant pressure" (Poynting 1903) 
in the particle's own frame of reference, or, "the process of absorption 
and re-emission produces no net force on a particle when one chooses to work 
with a stationary frame referred to the particle" (Wyatt and Whipple 1950).
The process of interaction of the incoming electromagnetic radiation
with the spherical particle and the resulting
motion of the particle is called the Poynting-Robertson (P-R) effect.
It is being included on regular basis into the modelling of
orbital evolution of cosmic dust grains under the action of electromagnetic
radiation (e. g., Poynting 1903, Robertson 1937, Wyatt and Whipple 1950,
Dohnanyi 1978, Jackson and Zook 1989, Gustafson 1994,
Dermott {\it et al.} 1994, Reach {\it et al.} 1995, Quinn 2005,
Gr\H{u}n 2007, Sykes 2007, Kr\H{u}gel 2008). The paper deals with an
inconsistency between Mie's solution of Maxwell's equations, relativity
theory and statements on the Poynting-Robertson effect.

\section{Condition for the P-R effect: current status}
Consider a spherical particle in its rest frame, and a beam of parallel
photons striking upon it. If the flux density of radiation energy
(energy flow through unit area perpendicular to the ray per unit time)
is $S$ and the geometric cross section of the particle
is $A$ ($\pi$ $\times$ the radius of the particle squared), then the incident
energy and momentum per unit time are
\begin{eqnarray}\label{1}
E_{i} &=& S~A ~,
\nonumber \\
\vec{p}_{i} &=& \frac{E_{i}}{c} ~ \vec{e} ~,
\end{eqnarray}
where $c$ is the speed of light and the unit vector $\vec{e}$ defines the
direction of the incoming photons (see also, e. g.,
pp. 5 and 11 in Burns {\it et al.} 1979). The condition
for the P-R effect, as generally accepted, relates
the total outgoing and incoming energies and momenta:
\begin{eqnarray}\label{2}
E_{o} &=& E_{i}  ~,
\nonumber \\
\vec{p}_{o} &=& \left ( 1 ~-~ Q_{pr} \right ) ~
		  \vec{p}_{i}
\end{eqnarray}
(see also, e. g., p. 5 and Eq. (10) on p. 10 in Burns {\it et al.} 1979),
where $Q_{pr}$ is the dimensionless efficiency factor for radiation pressure
(defined as the ratio $C_{pr} / A$, where $C_{pr}$ is cross section for the
radiation pressure) given by optical properties of the spherical
particle and wavelength(s) of the incoming radiation.
The first of Eqs. (2) states that the outgoing and incoming energies
are equal, which corresponds to the conservation of mass of the particle.
The rates of change of energy and momentum of the particle due to the
interaction with electromagnetic radiation are
\begin{eqnarray}\label{3}
\frac{d ~E}{d~ \tau} &=& E_{i}	~-~ E_{o} = 0 ~,
\nonumber \\
\frac{d ~\vec{p}}{d~ \tau} &=& \vec{p}_{i} ~-~ \vec{p}_{o} ~,
\end{eqnarray}
where $\tau$ is the proper time measured in the frame of reference of the
particle.

Eqs. (2) are in agreement with the statements published
since the time of Poynting (1903). The case of perfectly absorbing particle
corresponds to $Q_{pr}$ $=$ 1. Eqs. (2) yield $\vec{p}_{o}$ $=$ 0 for
$Q_{pr}$ $=$ 1 and this is equivalent to the statement that the process of
re-emission produces no net force on a particle in the
proper frame of reference of the particle, provided re-emission is the only
mechanism producing outgoing radiation (Poynting 1903, Robertson 1937,
Wyatt and Whipple 1950, and subsequent papers and books).

\section{Application of relativity theory}
We know, according to Einstein and Minkowski, that from the quantities
$E_{i}$, $\vec{p}_{i}$ and $E_{o}$, $\vec{p}_{o}$ we can compose
two four-vectors
\begin{eqnarray}\label{4}
p_{i}^{\mu} &=& \left ( \frac{E_{i}}{c}, \vec{p}_{i} \right ) =
		\left ( \frac{E_{i}}{c}, \frac{E_{i}}{c} ~\vec{e} \right ) ~,
\nonumber \\
p_{o}^{\mu} &=& \left ( \frac{E_{o}}{c}, \vec{p}_{o} \right ) =
		\left ( \frac{E_{i}}{c}, \frac{E_{i}}{c} ~
		\left ( 1 ~-~ Q_{pr} \right ) ~\vec{e} \right ) ~.
\end{eqnarray}
In these equations also Eqs. (1) and (2) have been used. From the
four-vectors $p_{i}^{\mu}$ and $p_{o}^{\mu}$ we can construct invariants
$M_{i}$ and $M_{o}$, with the physical meaning of the increase of mass
of the incoming and outgoing radiation per unit time. We have
\begin{eqnarray}\label{5}
M_{i} &=& \frac{1}{c} ~ \sqrt{p_{i}^{\mu} ~ p_{i ~ \mu}} \equiv
	  \frac{1}{c} ~ \sqrt{\left ( \frac{E_{i}}{c} \right ) ^{2} ~-~
	  \vec{p}_{i} \cdot \vec{p}_{i}} =
\nonumber \\
&=&  \frac{E_{i}}{c^{2}} ~ \sqrt{ 1 ~-~ \vec{e} \cdot \vec{e}}	= 0
\end{eqnarray}
and
\begin{eqnarray}\label{6}
M_{o} &=& \frac{1}{c} ~ \sqrt{p_{o}^{\mu} ~ p_{o ~ \mu}} \equiv
	  \frac{1}{c} ~ \sqrt{\left ( \frac{E_{o}}{c} \right ) ^{2} ~-~
	  \vec{p}_{o} \cdot \vec{p}_{o}} =
\nonumber \\
&=&  \frac{E_{i}}{c^{2}} ~ \sqrt{1 ~-~
     \left ( 1 ~-~ Q_{pr} \right ) ^{2}} ~.
\end{eqnarray}
Eq. (6) yields a nontrivial result:
1 $-$ $\left ( 1 ~-~ Q_{pr} \right ) ^{2}$ $=$
$Q_{pr}$ $\left ( 2 ~-~ Q_{pr} \right )$ $\ge$ 0, or,
\begin{equation}\label{7}
0 \le Q_{pr} \le 2 ~.
\end{equation}

\section{Application of Mie's theory}
Mie's theory is a solution of Maxwell's equations for electricity and magnetism.
The solution holds for the case when an incoming electromagnetic radiation
interacts with a spherical particle. Mie's theory offers also the values of
$Q_{pr}$, the fundamental quantity for the P-R effect.
What are the values of $Q_{pr}$? For our purposes it is important that they can
be larger than 2 [e. g., van de Hulst (1981) presents also values
larger than 2.5 in Table 13 on p. 161].

\section{Inconsistency in physics}
We have got a problem. Relativity theory states that $0 \le Q_{pr} \le 2$
(see Eq. 7), but Mie's theory offers also values $Q_{pr} > 2$!

Relativity theory was motivated by Maxwell's equations. However, Mie's theory
is a special solution of the Maxwell's equations.
Why an inconsistency exists between these two approaches?
Why the inconsistency exists 100 years?

The conclusion is that physicists and astrophysicists sometimes use
also theories which yield inconsistent results. But instead of being
disappointed, an optimistic view is possible: "How wonderful that we have met
with paradox. Now we have some hope of making progress." (Niels Bohr).

\section{Fundamental condition for the P-R effect: correct statement}
In order to solve the paradox, we have to realize that the statement
"the process of re-emission produces no net force
on a perfectly absorbing spherical particle when one chooses to work with
a stationary frame referred to the particle" ($Q_{pr}$ $=$ 1 in Eq. 2)
concerns only a part of the outgoing radiation. In the complete
description, one has to take into account also diffraction
(small angle scattering) of the light. As a matter of fact, there exists
"extinction paradox" (van de Hulst 1981, p. 107) according to which
diffracted light plays a non-negligible role in treating
the incoming and outgoing radiation, even for particles that are large
in comparison with the wavelength of the interacting light; thus,
diffraction must be included into considerations even in the situation
in which one normally gets along with the geometrical optics approximation.
We know that the diffracted light gives a zero contribution to the radiation
pressure of large spheres (van de Hulst 1981, p. 225), but diffraction cannot
be neglected in a separate treatement of the incoming and outgoing radiation.

The correct physics gives the following result for a large perfectly absorbing
spherical particle: the dimensionless efficiency factors for absorption and
scattering are $Q_{abs}$ $=$ 1 (this result is used in the P-R effect:
$E_{i}$ $=$ $S A Q_{abs}$ $=$ $S A$ in Eq. (1)) and $Q_{sca}$ $=$ 1 (due to
the diffraction). These two factors  sum up
into the efficiency factor of extinction $Q_{ext}$ $=$ $Q_{abs}$ $+$ $Q_{sca}$
(cross sections can be obtained by multiplication with geometrical cross
section $A$). The correct result for the large perfectly absorbing spherical
particle, with the effect of diffraction taken into acount, is $Q_{ext}$ $=$ 2.
The effect of diffraction is equally important as the effect of absorption.
As a consequence, the conventional condition for the P-R effect $E_{i}$ $=$
$S A$ (Eq. (1)) has to be replaced by the physical condition $E_{i}$ $=$
$S A Q_{ext}$. Thus, we have the following condition for the incoming radiation
\begin{eqnarray}\label{8}
E_{i} &=& S~A~Q_{ext} ~,
\nonumber \\
\vec{p}_{i} &=& \frac{E_{i}}{c} ~ \vec{e},
\end{eqnarray}
instead of Eq. (1).
Moreover, the second condition for the P-R effect cannot be of the form (2).
It can be easily seen from Eqs. (1)-(3) and Eq. (8) that Eqs. (3) do not change
if
\begin{eqnarray}\label{9}
E_{o} &=& E_{i}  ~,
\nonumber \\
\vec{p}_{o} &=& \left ( 1 ~-~ \frac{Q_{pr}}{Q_{ext}} \right ) ~
		  \vec{p}_{i} ~.
\end{eqnarray}

Eqs. (8) and (9) are the conditions under which the P-R effect holds.
The conditions are formulated in the proper frame of reference of the particle.
The case of a large perfectly absorbing sphere corresponds to $\vec{p}_{o}$ $=$
$\vec{p}_{i}$  $/$ 2 and not to $\vec{p}_{o}$ $=$ 0 as it has been
conventionally stated.

The expression for the four-vector $p_{o}^{\mu}$,
\begin{eqnarray}\label{10}
p_{o}^{\mu} &=& \left ( \frac{E_{o}}{c}, \vec{p}_{o} \right ) =
		\left ( \frac{E_{i}}{c}, \frac{E_{i}}{c} ~
		\left ( 1 ~-~ \frac{Q_{pr}}{Q_{ext}} \right ) ~
		\vec{e} \right ) ~,
\end{eqnarray}
yields for the increase of mass of the outgoing radiation per unit time
\begin{eqnarray}\label{11}
M_{o} &=& \frac{1}{c} ~ \sqrt{p_{o}^{\mu} ~ p_{o ~ \mu}}
\nonumber \\
&=&  \frac{E_{i}}{c^{2}} ~ \sqrt{1 ~-~
     \left ( 1 ~-~ \frac{Q_{pr}}{Q_{ext}} \right ) ^{2}} ~.
\end{eqnarray}
The non-negativity of the expression under the square root symbol is
equivalent to
\begin{equation}\label{12}
0 \le \frac{Q_{pr}}{Q_{ext}} \le 2 ~.
\end{equation}
Physical condition represented by Eq. (12) differs from the condition
presented in Eq. (7) and is consistent with Mie's solution of
Maxwell's equations.

\section{Conclusion}
A paradox coming from the application of relativity theory and Mie's solution
of Maxwell's equations to the Poynting-Robertson effect is presented.
Solution of the paradox resides in the fact that diffraction plays a
non-negligible role in the process of interaction between the incoming
radiation and the spherical particle. As a consequence, in the case of a large
perfectly absorbing spherical particle there holds the condition 
$\vec{p}_{o}$ $=$ $\vec{p}_{i}$ $/$ 2
for the outgoing and incoming momenta per unit time,
and not the conventional condition $\vec{p}_{o}$ $=$ 0 obtained by the
neglection of diffraction. Physically correct relations defining
the P-R effect are given by Eqs. (8) and (9).

\begin{acknowledgements}
The author is indebted to V. Balek for his valuable comments.
This work was supported by the Scientific Grant Agency VEGA, Slovakia,
grant No. 1/3074/06.
\end{acknowledgements}

\end{document}